\documentclass[conference]{IEEEtran}
\IEEEoverridecommandlockouts
\usepackage{cite}
\usepackage{amsmath,amssymb,amsfonts}
\usepackage{algorithmic}
\usepackage{graphicx}
\usepackage{textcomp}
\usepackage{xcolor}
\usepackage{url}
\usepackage{breakurl}
\usepackage[breaklinks]{hyperref}
\usepackage[nolist]{acronym}
\def\BibTeX{{\rm B\kern-.05em{\sc i\kern-.025em b}\kern-.08em
    T\kern-.1667em\lower.7ex\hbox{E}\kern-.125emX}}
    
\begin{document}

\newacro{EUD}[EUD]{End-user Development}
\newacro{EUP}[EUP]{End User Programming}
\newacro{EUSE}[EUSE]{End User Software Engineering}
\newacro{UML}[UML]{Unified Modeling Language}
\newacro{DSL}[DSL]{Domain Specific Language}
\newacro{HTML}[HTML]{Hypertext Markup Language}
\newacro{CSS}[CSS]{Cascading Style Sheets}
\newacro{VR}[VR]{Virtual Reality}
\newacro{AR}[AR]{Augmented Reality}
\newacro{MIME}[MIME]{Multipurpose Internet Mail Extensions}
\newacro{MR}[MR]{Mixed Reality}
\newacro{HMD}[HMD]{Head Mounted Display}
\newacro{HMDs}[HMDs]{Head Mounted Displays}
\newacro{ECS}[ECS]{Entity-Component-System}
\newacro{XRT}[XRT]{X-Reality Toolkit}
\newacro{IDE}[IDE]{Integrated Development Environment}
\newacroplural{IDEs}[IDE]{Integrated Development Environments}
\newacro{GREP}[GREP]{Game Rules scEnario Platform}
\newacro{GREM}[GREM]{Game Rules scEnario Model}
\newacro{glTF}[glTF]{GL Transmission Format}
\newacroplural{HMD}[HMDs]{Head Mounted Displays}
\newacro{3D}[3D]{three-dimensional}
\newacro{2D}[2D]{two-dimensional}
\newacro{SUS}[SUS]{System Usability Scale}
\newacro{URL}[URL]{Uniform Resource Locator}
\newacroplural{URL}[URLs]{Uniform Resource Locators}
\newacro{VRLE}[VRLE]{Virtual Reality Learning Environment}
\newacroplural{VRLE}[VRLEs]{Virtual Reality Learning Environments}
\newacro{CVE}[CVE]{Collaborative Virtual Environment}
\newacroplural{CVE}[CVEs]{Collaborative Virtual Environments}

\title{VREUD - An End-User Development Tool to Simplify the Creation of Interactive VR Scenes
}

\author{\IEEEauthorblockN{Enes Yigitbas, Jonas Klauke, Sebastian Gottschalk, Gregor Engels}
\IEEEauthorblockA{\textit{Paderborn University}, Germany \\
firstname.lastname@upb.de}}


\maketitle


\begin{abstract}
	Recent advances in Virtual Reality (VR) technology and the increased availability of VR-equipped devices enable a wide range of consumer-oriented applications. For novice developers, however, creating interactive scenes for VR applications is a complex and cumbersome task that requires high technical knowledge which is often missing. This hinders the potential of enabling novices to create, modify, and execute their own interactive VR scenes. Although recent authoring tools for interactive VR scenes are promising, most of them focus on expert professionals as the target group and neglect the novices with low programming knowledge. To lower the entry barrier, we provide an open-source web-based End-User Development (EUD) tool, called VREUD, that supports the rapid construction and execution of interactive VR scenes. Concerning construction, VREUD enables the specification of the VR scene including interactions and tasks. Furthermore, VREUD supports the execution and immersive experience of the created interactive VR scenes on VR head-mounted displays. Based on a user study, we have analyzed the effectiveness, efficiency, and user satisfaction of VREUD which shows promising results to empower novices in creating their interactive VR scenes. 
\end{abstract}


\section{Introduction}
Recent advances in \ac{VR} technology and the increased availability of VR-equipped devices slowly shift the focus away from the entertainment area to new application areas in training~\cite{DBLP:conf/vrst/YigitbasJSE20}, robotics \cite{DBLP:journals/corr/abs-2103-10804}, education~\cite{DBLP:conf/mc/YigitbasTE20}, or healthcare~\cite{DBLP:conf/mc/YigitbasHE19}. 
These wider application areas of \ac{VR} require, besides affordable devices, a usable process of authoring to reach the full potential \cite{PotentialVR}. Although there is a vast pool of authoring tools \cite{AuthoringGroups}, many of them presuppose technological background in programming or scripting \cite{UnityWebpage, UnrealWebpage, PolyVR} or knowledge in modeling languages \cite{SumerianWebpage, FlowMatic}. This results in a complex and cumbersome task to develop interactive \ac{VR} scenes for novices in \ac{VR} development, who have typically lower technical knowledge.

One solution to this problem is to apply \ac{EUD} methods and techniques which allow novices to create or modify software artifacts \cite{EUDgrundlage}. Furthermore, modeling languages are used in authoring tools to abstract the complexity of the development, however, the entry barrier to learning the language could scare away novices. As a consequence, we believe that the development of an interactive \ac{VR} scene should require a minimal entry barrier to reach a maximum of novices. To reach this goal, we have chosen the EUD methods component-based and wizard-based development. The component-based development allows an efficient construction by reusing and configuring of provided components. Furthermore, we apply wizard-based development to lead the novice in a step-by-step manner through difficult sections in the development. These chosen \ac{EUD} methods are implemented in an open-source web-based authoring tool called VREUD. It addresses the challenges in the development of an interactive \ac{VR} scene by supporting the novice in the construction of the scene. Furthermore, it supports the specification of interactions which will be performed by the user of the VR scene, and the definition of tasks to guide the VR user through the scene. Since novices usually prefer to learn by exploring or trial and error mechanism \cite{noManual}, VREUD provides rapid construction and execution that enables the novices to prototype the developed interactive \ac{VR} scene at each step in the development. Consequently, VREUD enables the novice to construct and execute the developed web-based interactive \ac{VR} scene.
VREUD is inspected by a usability evaluation to analyze its effectiveness, efficiency, and user satisfaction. 15 participants with mixed backgrounds in \ac{VR} development have been invited to a user study to evaluate if VREUD supports a simple development of interactive VR scenes with a low entry barrier. The results of the usability evaluation suggest that VREUD enables novices to successfully create their interactive VR scenes and that VREUD is easy to learn and decreases the entry barrier in the development. 

Our contributions are as following: A technique to abstract the development of interactions in an interactive \ac{VR} scene by using a component-based and wizard-based construction, an open-source web-based authoring tool to develop web-based interactive \ac{VR} scenes with a low entry barrier that can be used or extended by other researchers, and a usability evaluation inspecting the effectiveness, efficiency, and user satisfaction of the introduced authoring tool as well as a comparison between the results of users with no and high experience in \ac{VR} development.

The rest of the paper is structured as follows. In Section 2, we present and discuss the related work. In Section 3, we described the conceptual solution of VREUD. In Section 4, we show the details of the implementation of VREUD. In Section 5, we present the main results of the usability evaluation and discuss its results. In Section 6, we conclude the paper and give an outlook for future work.

\section{Related Work}

Augmented Reality (AR) and Virtual Reality (VR) have been a topic of intense research in the last decades. In the past few years, massive advances in affordable consumer hardware and accessible software frameworks are now bringing AR and VR to the masses.  AR enables the augmentation of real-world physical objects with virtual elements and has been already applied for different aspects such as robot programming \cite{DBLP:journals/corr/abs-2106-07944}, product configuration (e.g., \cite{DBLP:conf/hcse/GottschalkYSE20}, \cite{DBLP:conf/hcse/GottschalkYSE20a}), prototyping \cite{DBLP:conf/hcse/JovanovikjY0E20}, planning and measurements \cite{DBLP:conf/eics/EnesScaffolding} or for realizing smart interfaces (e.g., \cite{DBLP:conf/eics/KringsYJ0E20}, \cite{DBLP:conf/interact/YigitbasJ0E19}). In contrast to AR, VR interfaces support the interaction in an immersive computer generated 3D world and have been used in different application domains as motivated already in the introduction. As the creation of interactive VR scenes is our main focus, in the following, we draw on prior research into \textit{\ac{VR} Authoring Tools} and \textit{End-User Development}.

\subsection{VR Authoring Tools}
\ac{VR} authoring tools enable the user to develop interactive \ac{VR} scenes. They provide the user with the ability to position virtual objects in a scene and methods to define the interactions between these objects. The game engines from Unity \cite{UnityWebpage} and Unreal \cite{UnrealWebpage} are currently the most popular tools to develop interactive \ac{VR} scenes \cite{UnityUnrealPopular}. The area of WebVR, which enables \ac{VR} on web pages, has produced frameworks to develop interactive \ac{VR} scenes as web applications, for example, Threejs \cite{ThreejsWebpage} and A-Frame \cite{AFrameWebpage}. The problem with these introduced tools is that they require programming languages like C\# and JavaScript that creates an entry barrier for novices. Arising from this problem, many authoring tools provide an easier development of scenes to novices \cite{SpokeWebpage, SumerianWebpage, EUDGames-GREP, XRT, PolyVR}. In Spoke \cite{SpokeWebpage}, the user designs social meeting scenes for Hubs\cite{HubsWebpage}. Hubs is a web-based VR social meeting platform. Spoke simplifies the development of the scene to a combination of elements given to the user. These elements can be modified and configured in their appearance. The design of user interactions is not possible. PolyVR \cite{PolyVR} supports the design of interactions by scripting. The downside of this approach is that the novice has to learn the modeling language. Sumerian \cite{SumerianWebpage} abstracts interactions with statecharts. This decreases the entry barrier, but the novices have to be familiar with statecharts. \ac{GREP} \cite{EUDGames-GREP} supports novices in the development of \ac{VR} games. It decreases the complexity of the construction by using smaller games that are defined by a given archetype. These smaller games are configured and combined to a resulting more complex game. This enables the design of user guidance in the developed VR scene but the user interactions are only indirectly designed by the chosen archetypes. This shows a lack of an authoring tool that supports novices in the design of the scene, interactions, and tasks. Immersive authoring tools enable the novice to create and modify the interactive \ac{VR} scene directly inside the 3D scene instead of the creation and modification of a 2D abstraction of the scene in non-immersive authoring tools. Another advantage is that \ac{HMDs} provide the novices with natural interactions. VR GREP \cite{ImmersiveGREP} simplifies the development of the scene by supplying the novice with a set of objects to place in it. The novice can manipulate them directly with natural interactions. However, in VR GREP, the user is not capable to design interactions or tasks in an immersive way. This is solved in FlowMatic \cite{FlowMatic} by the usage of functional reactive programming \cite{FRP}. The approach provides the novice with a way to create interactions in a visual model by connecting nodes, which are objects inside the scene, and operations, that transform them. But similar to the usage of statecharts, this presupposes learning for novices. The results of \cite{ImmersiveBenefitsGREP} show that immersive authoring benefits the authoring of \ac{VR}, however, the current natural interactions in \ac{VR} lack the accuracy of a desktop solution. Additionally, a \ac{HMD} is required to construct the scene and long developing sessions are tiring for the user. As a consequence, a desktop authoring tool holds still benefits compared to an immersive authoring tool. The immersive and non-immersive way of authoring VR scenes is combined in SAVEace \cite{immersiveDesktop} to benefit from both approaches. In summary, we can conclude that existing immersive and non-immersive VR authoring tools still require a high degree of programming knowledge which hinders the rapid creation of interactive VR scenes for novices. 

\subsection{End-User Development}
The related work of \ac{EUD} is focused on component-based development and wizard-based development. 

In component-based development, the software is described by a connection of components. These components have an embedded functionality and a defined interface to use them. These interfaces are used to connect them. This abstracts the development. With the help of visual or textual interfaces, the connection of components can be further abstracted to decrease the entry barrier for novices. Another advantage is the flexible construction of software since components can be added easily to the component network or existing components can be switched with other components. 
In \cite{componentChemie}, component-based development is combined with a domain-specific language to enable chemistry teachers the development of virtual chemistry experiments. This resulted in a fast and easy construction. Furthermore, in \cite{DBLP:conf/eics/YigitbasJJKAE19, DBLP:journals/sosym/YigitbasJBSE20, DBLP:journals/pacmhci/YigitbasHRASE19}, component-based development has been applied to support the engineering of adaptive user interfaces. Cicero Designer \cite{ComponentMuseum} enables novices the development of multi-device museum guides. XRT \cite{XRT} enables novices the design of cross-reality interactions by a connection of predefined components. This shows that component-based development is an often-used approach to supply novices with methods to develop software themselves. In wizard-based Development, the user is guided through the development process with a wizard. This requires that the performed task can be split into smaller steps. As a consequence, the user is focused on one step instead of all steps at the same time. This approach is combined with other methods in \cite{wizardSmartHome} to customize the visualization of data in smart homes and in \cite{wizardEGov} to enable novices to develop e-government services. Altogether we can observe that existing EUD methods are already used in various application domains to ease the development of different application types. Inspired by these approaches, our goal is to provide an end-user development tool to simplify the creation of interactive VR scenes.   

\section{Solution Overview}
The goal of VREUD is to decrease the complexity of the development of interactive \ac{VR} scenes by supporting novices with a way to create interactive VR scenes themselves. Component-based and wizard-based development are chosen to minimize the learning effort for novices to guarantee a low entry barrier. Consequently, the interactive \ac{VR} scene is built completely by the combination of components, and complex development processes are split by the usage of step-by-step wizards. In our solution approach, the development of interactive VR scenes consists of the following steps: \textit{Construction of the Scene}, \textit{Construction of Interactions}, and \textit{Construction of Tasks}. 

\subsection{Construction of the Scene}
\begin{figure}[htbp]
\centerline{\includegraphics[scale=0.6]{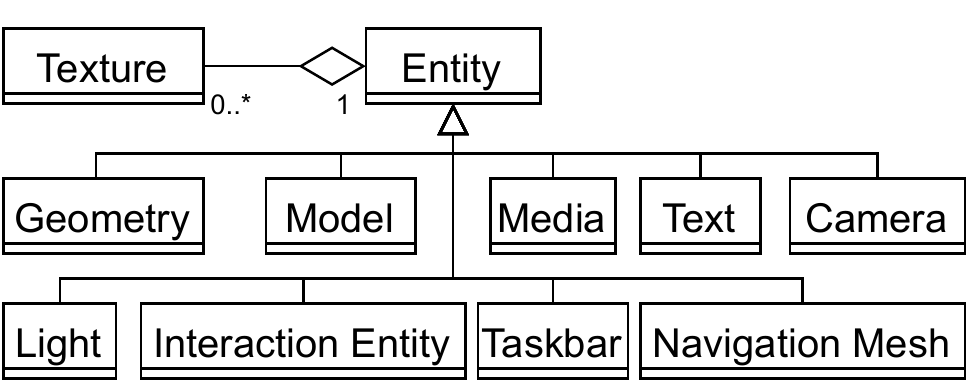}}
\caption{Class diagram of the entities to describe the scene in VREUD}
\label{entities}
\end{figure}

The construction of the scene supports novices in the development of the visual appearance of their interactive VR scene. The development is abstracted by a combination of entities to build the VR scene. These entities describe 3D virtual objects that define the appearance of the VR scene. To provide a comprehensive design of VR scenes the entities have sub-classes to fit different needs of the specific virtual object. These types are shown in the class diagram in Fig. \ref{entities}. For example, the entities can describe the \textit{Geometry} primitives like a box or cylinder, 3D \textit{Models} which can be added from external sources, external \textit{Media} like pictures, videos, and PDF documents, \textit{Interaction Entities} like a button, a counter, or a pressure plate which provide new interactions methods, \textit{Taskbars} which show tasks to the \ac{VR} user and a \textit{Navigation Mesh} which limits the \ac{VR} user in the navigation by defining the area the \ac{VR} user can move onto.

Every entity contains parameters, that define the representation and behavior of the virtual object. For example, a 3D box has representative parameters depth, height, and width to configure the shape of the box and color, or texture to set the appearance. The parameters position and rotation define the location of the entity in the scene and behavioral parameters configure, for example, the shadow and physics of the entity.

The novices are supported in the construction by visualizing the interactive VR scene in the development. This ensures direct feedback of added entities to the scene and modified representative parameters of the entities inside the scene. This enables the novice to spot mistakes immediately. The modification of parameters is either performed in a form-based representation of all parameters and the associated values, or by a direct manipulation interface that utilizes the visualization of the interactive VR scene to change position, rotation, and scale of the entity directly inside the shown interactive VR scene.    

\subsection{Construction of Interactions}
To transform the developed static VR scene into an interactive \ac{VR} scene, interactions have to be added. Interactions describe modifications of the entities inside the VR scene. They enable the VR user to interact with the entities inside the interactive VR scene. The developed interactions are event-based. Every interaction describes only an atomic modification of a specific entity and is triggered by a single event. As a consequence, the resulting interaction is not powerful, but interactions can be combined to create more expressive interactions. However, this restriction allows defining interactions by components that are configured by a simple selection of given elements. An interaction is defined by the following elements:
\textbf{An Event}, that describes the trigger of an interaction. If the event occurs the designed interaction is performed. These events can be either triggered by the \ac{VR} user or other entities in the scene. For example, when the \ac{VR} user touches the entity and presses the trigger button on the controller. 
\textbf{An Effect}, that defines the modification of an entity in the scene. For example, an effect can change the color of an entity.
\textbf{Parameters}, which might be needed to perform the effect. For example, the color blue to change the entity to blue.
\textbf{Entities}, that define the context of the interaction. Every interaction is defined by two entities. These entities can be the same entity or two different entities. The source entity defines the source of the event and the target entity sets the target of the effect.
\textbf{Conditions}, which can be added to an interaction. Conditions define when an interaction can be performed by a \ac{VR} user in the scene. All applied conditions have to be fulfilled to perform the effect of the interaction. For example, the interaction can only be performed if an entity is placed in a specific location.

\begin{figure}[htbp]
\centerline{\includegraphics[scale=0.45]{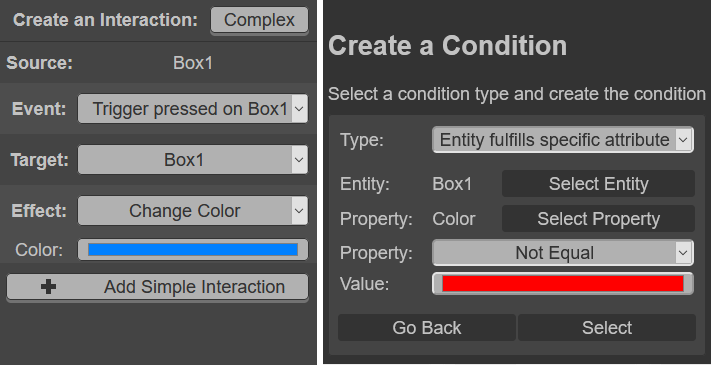}}
\caption{The interface of the interaction construction. The current configuration results in an interaction that changes the color of Box1 to blue if the controller touches Box1 and the trigger is pressed. The complex button opens a wizard to add conditions. The configured condition checks if Box1 has not the color red before performing the interaction.}
\label{specifaction}
\end{figure}

A simple interaction can be performed by the \ac{VR} user at every time. As a result, the interaction contains no conditions. This allows the usage of an interface that selects only a source entity, an event, a target entity, an effect, and parameters needed for the effect which is shown on the left side of Fig. \ref{specifaction}. Since the event is triggered on the source entity and the effect is performed on the target entity, the chosen source entity defines the set of possible events and the chosen target entity defines the set of possible effects. The interface adapts the chosen event and effect to changes in the selected source and target entity. This guarantees a correct construction of the interaction since the novice can not choose an event or effect which is not possible for the chosen source or target entity.

\begin{figure}[htbp]
\centerline{\includegraphics[scale=0.38]{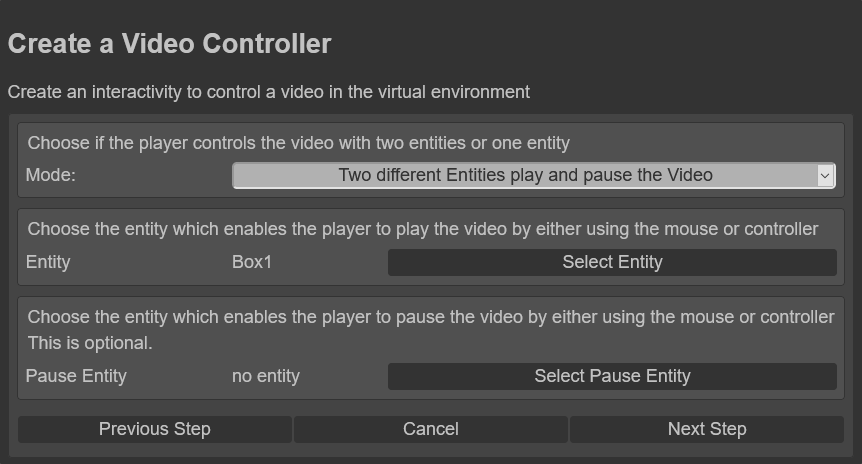}}
\caption{The pattern dialog to create interactivity around a video. The user creates interactions just by the selection of entities. Box1 is configured as entity to play the video by pressing on it.}
\label{patternDialog}
\end{figure}

A complex interaction is an interaction that can not always be performed by the \ac{VR} user. Consequently, conditions are applied to the construction of interactions. To not overload the interface, a button is added to the interaction interface shown on the left side of Fig. \ref{specifaction} to open a wizard that performs the construction of the complex interaction. The setup of a condition is shown on the right side of Fig. \ref{specifaction}. A condition is configured by a condition type and parameters associated with the chosen type. This enables the usage of a component to check the specific condition that is configured by a simple selection. For example, the condition type \textit{Entity fulfills a specific attribute} creates a condition that checks if a chosen attribute of a selected entity fulfills a configured expression. As a result, the complex interaction can only be performed by the \ac{VR} user if the attribute of the entity fulfills the expression. The conditions can be combined to define more complex conditions. The interface of the condition construction adapts to already created conditions to prevent conditions that conflict with each other. The previously introduced interfaces produce interactions for atomic actions that require an understanding of the events and effects in the scene. To further decrease the cognitive effort required to perform the construction, interactivity patterns are supported in VREUD. These patterns are components in the editor which use a wizard to configure the pattern. The components have defined input and output to ensure the easy extension of new patterns. The patterns can result in a set of added entities, interactions, tasks, and changed parameters of existing elements. As a result, these patterns can perform a set of construction steps in one step. Consequently, the abstraction level is higher. Fig. \ref{patternDialog} shows one step in the wizard of the construction of the interactivity associated with a video in the scene. In the pattern, the novice has to only select entities, which, in this example dialog, start and pause the video. As a result, the novice has not created both interactions manually which saves her time in the construction.

\subsection{Construction of Tasks}
\begin{figure}[htbp]
\centerline{\includegraphics[scale=0.43]{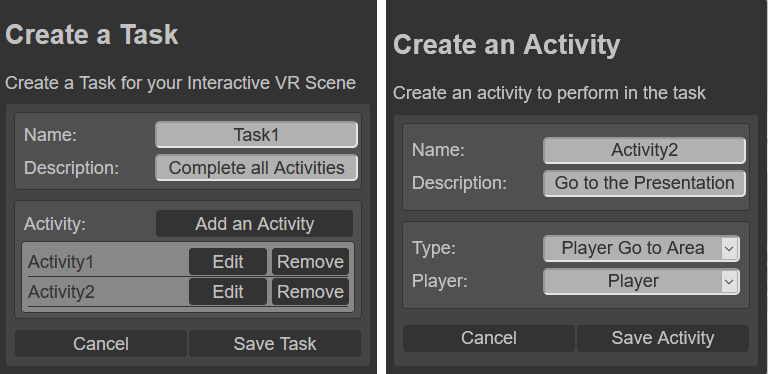}}
\caption{The dialog to create a task and the dialog to create an activity. The task consists of two activities and the shown activity tracks if the \ac{VR} user has entered a specific location.}
\label{taskDialog}
\end{figure}

\begin{figure}[htbp]
\centerline{\includegraphics[scale=0.3]{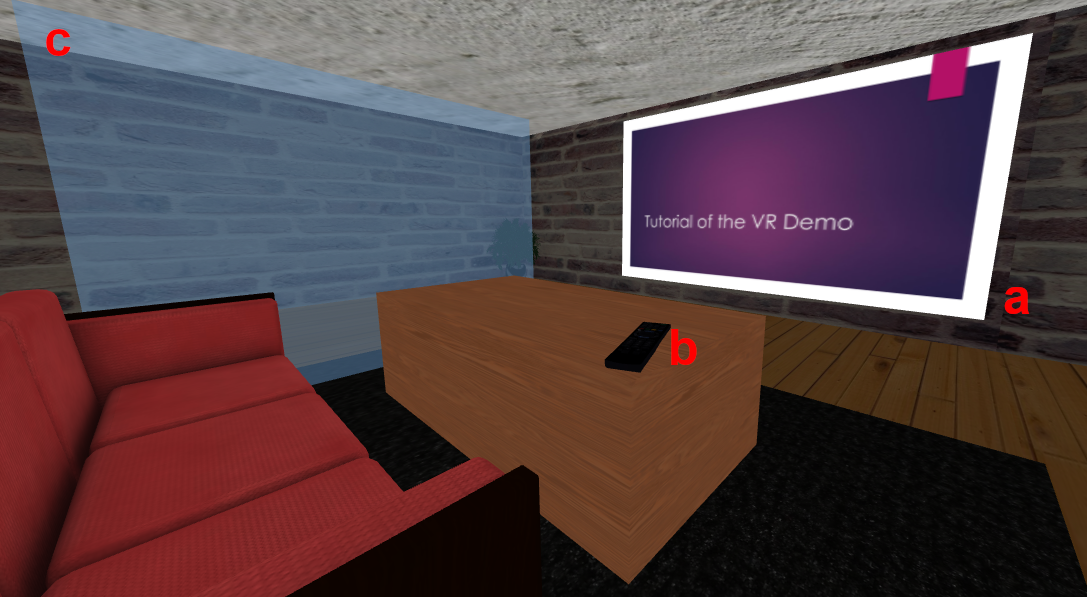}}
\caption{An interactive \ac{VR} scene developed with VREUD. It is a living room that shows a presentation (a) to the user. An interaction enables the user to open the next page by pressing on the remote (b). To complete the task, the user has to go to the area (c).}
\label{scene}
\end{figure}

The construction of tasks supports novices in the development of tasks to guide the VR user through their developed interactive VR scene. Tasks define actions the \ac{VR} user has to perform in the interactive \ac{VR} scene. The construction of tasks is inspired by the approach of \ac{GREP}\cite{EUDGames-GREP}, which resulted in an easy-to-use and understandable construction of education games. Similar to \ac{GREP}, a task is defined by a set of smaller tasks. These smaller tasks will be called activities. An activity is configured by an activity type that defines the atomic action the VR user has to perform. This action is tracked by the activity in the interactive VR scene. This allows encapsulating the activity in a component that is configured to the specific needs of the novice. Since the VR user has to be aware of the tasks inside the interactive VR scene, the tasks and activities have parameters to configure the names and the descriptions that are shown inside the interactive VR scene. The construction uses two dialogs to create the task and the activities so that the novice only needs to focus on the current task or activity. The two dialogs are shown in Fig. \ref{taskDialog}. Since some activity types are location dependable, an area has to be defined. These areas are spawned after the task construction and they are set up similar to the entities in the construction of the scene. These areas are also used to decrease the issue to describe the area in the activity description by showing the transparent area directly in the interactive VR scene to the \ac{VR} user. An area is shown in Fig. \ref{scene}. To not overload the scene, the area of a solved activity disappears. 

\section{Implementation}
\begin{figure}[htbp]
\centerline{\includegraphics[scale=0.52]{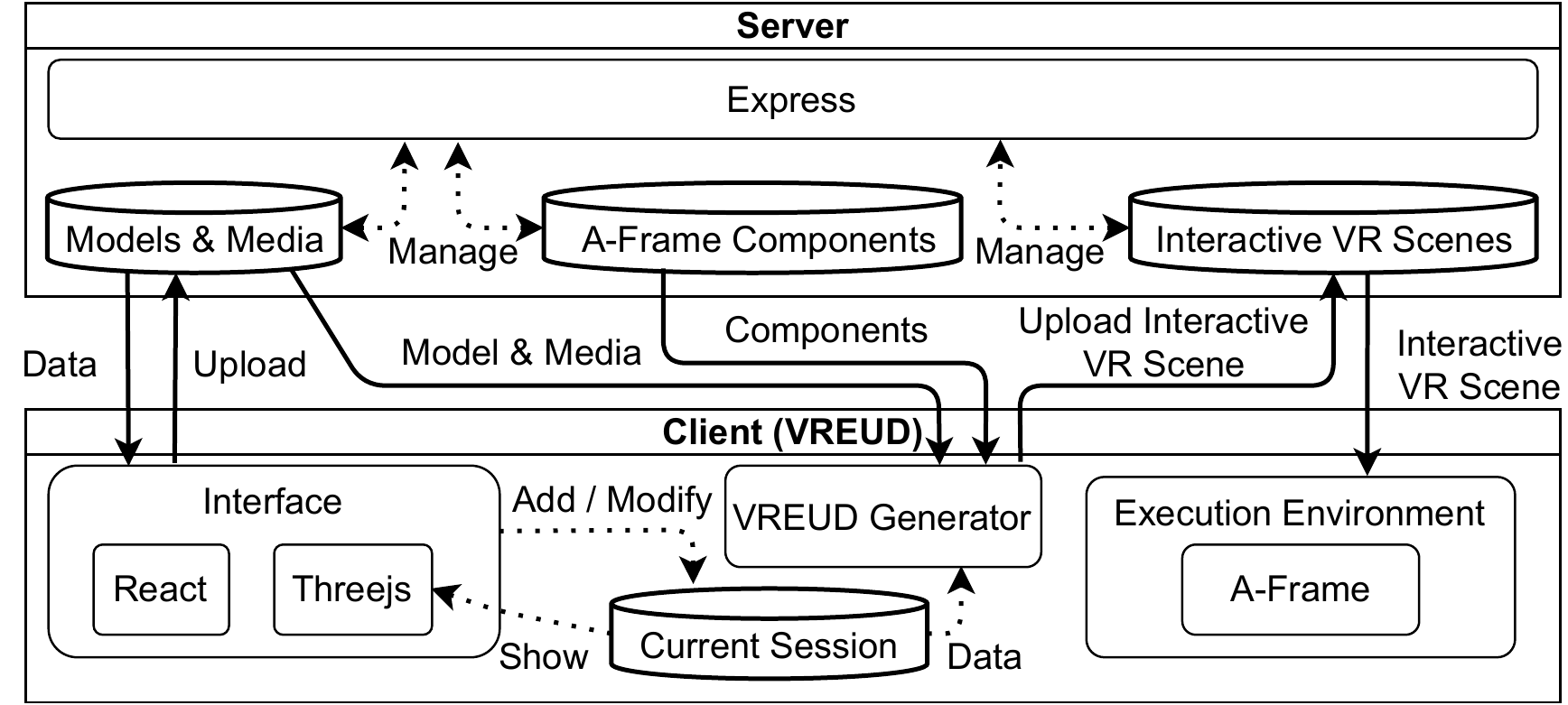}}
\caption{The system architecture of \textit{VREUD}}
\label{architecture}
\end{figure}

\begin{figure}[htbp]
\centerline{\includegraphics[scale=0.45]{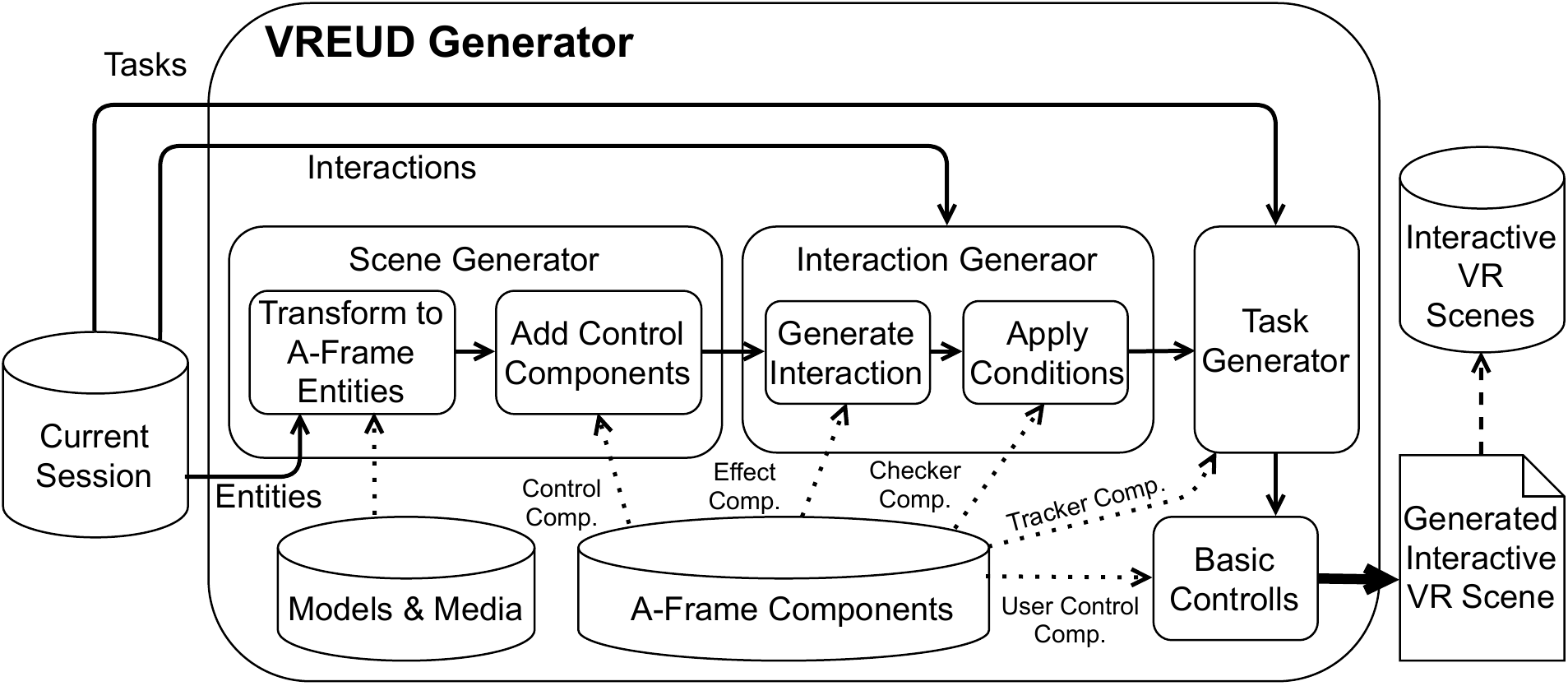}}
\caption{Architectural overview of the \textit{VREUD Generator}}
\label{generator}
\end{figure}

\begin{figure*}[htbp]
\centerline{\includegraphics[scale=0.4]{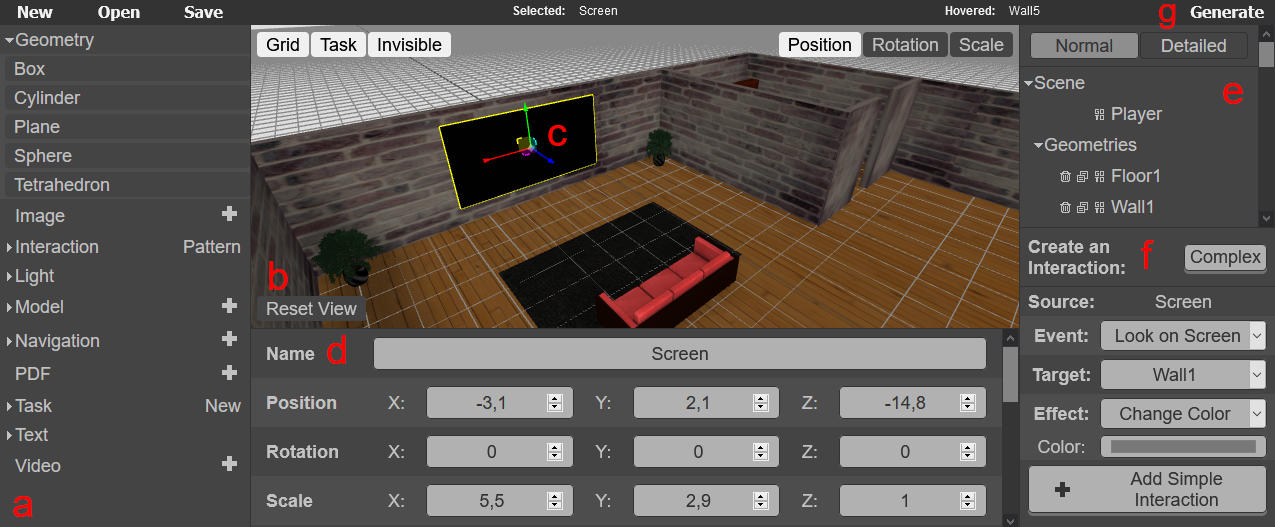}}
\caption{The interface of the VREUD. (a) lists all entities to instantiate in the scene, (b) visualizes the interactive VR scene, (c) shows the direct manipulation interface of the selected entity, (d) shows the form to modify parameters of the selected entity, (e) lists all entities, interactions, and tasks of the interactive \ac{VR} scene, (f) shows the interface to create interactions and (g) executes the developed interactive VR scene.}
\label{interface}
\end{figure*}

VREUD is an open-source web-based authoring tool\footnote{\url{https://github.com/VREUD/VREUD}}. VREUD has a server-side and a client-side. The system architecture is presented in Fig. \ref{architecture} and the architecture of the \textit{VREUD Generator}, that generates the interactive VR scenes, is shown in Fig. \ref{generator}. The server-side uses \textit{Express} \cite{Express} as a web server. The stored data on the server is divided into three repositories. \textit{Models and Media} stores added models and media by the novices, \textit{Interactive VR Scenes} saves the generated interactive VR scenes, and \textit{A-Frame Components} stores components that implement the developed interactions and required behavior of the entities since the generated interactive VR scene uses \textit{A-Frame}\cite{AFrameWebpage}. A-Frame is a web-based VR framework. The VREUD client-side utilizes \textit{React} \cite{ReactWebpage}, a library for building user interfaces, and \textit{Threejs} \cite{ThreejsWebpage}, a framework for web-based 3D scenes, to implement the component-based interface and to show the interactive VR scene directly in the interface. The component-based architecture enables easy extensions of new features in the interface. The interface is presented in Fig. \ref{interface}. The interactive VR scene is defined by objects instantiated from a class system implemented in JavaScript that provides objects for activities, entities, interactions, and tasks. These objects are stored in the \textit{Current Session}. VREUD uses these objects to generate the interactive \ac{VR} scene with the help of the \textit{VREUD Generator} component. The generated interactive \ac{VR} scene is translated in A-Frame. A-Frame builds on Threejs to describe VR scenes. Consequently, both representations of the interactive \ac{VR} scene do not conflict with each other, since both use Threejs.

The \textit{VREUD Generator} produces an interactive \ac{VR} scene implemented in A-Frame from a given set of objects stored in \textit{Current Session} on the client-side. Virtual objects in A-Frame are defined by an entity-component-system architecture. In it, virtual objects are described by entities containing components. These components define the appearance or the behavior of the entity. To implement the component-based approach of VREUD, the generator uses predefined components that are configured by the novices. These components are stored in \textit{A-Frame Components}. The architecture of the generator is shown in Fig. \ref{generator}. First of all, in the \textit{Scene Generator}, the entity is mapped to an \textit{A-Frame} entity to implement the appearance. If an external model or media is used, the generator maps the entity to the stored model or media in \textit{Models and Media}. Then \textit{Control Components} are added if the entity requires them. For example, a PDF document uses them to control the shown page. After this step, the scene is completely generated and interactions are added by the \textit{Interaction Generator}. These interactions are implemented by \textit{Effect Components} which are associated with the chosen effect in the interaction construction. The component is configured and applied to the source entity since the event has to be listened at this entity. The optional conditions are also implemented by \textit{Checker Components} which are also configured by the novice. The \textit{Task Generator} adds new invisible A-Frame entities to the scene to implement the tasks and activities. Activity entities use \textit{Tracker Components} to check the configured actions in the interactive \ac{VR} scene. If the activity uses an area, a 3D box is used as the body. The activities are connected to the tasks to inform them about the completion. Finally, basic user controls are applied to the scene by the usage of specific \textit{User Control Components}. The resulting interactive VR scene is stored in \textit{Interactive VR Scenes} on the server-side to provide access to the \ac{VR} user. 

In summary, VREUD can execute interactive VR scenes. To utilize this, the construction is designed to enable prototyping at each step to experience the current interactive \ac{VR} scene. This provides direct feedback of the developed interactive VR scene and the novice can explore the construction. Additionally, an interactive tutorial supports the novices in their first construction.

\section{Evaluation}
We conducted a usability evaluation to evaluate the effectiveness, efficiency, and user satisfaction of VREUD. To rate the effectiveness and efficiency, the participants had to solve assignments whose required time and successful completion are measured. The user satisfaction is measured by a questionnaire. The evaluation is performed to gain insights into the required entry barrier of VREUD and to assess the construction of scenes, interactions, and tasks. To estimate the usefulness of VREUD for novices, we have invited participants with mixed knowledge in \ac{VR} development to compare the results of \ac{VR} novices against \ac{VR} experts. We have decided to perform the usability evaluation remotely, due to the situation caused by COVID and the web-based architecture of VREUD that provides easy remote access.

\subsection{Participants and Procedure}
We recruited 15 volunteer participants for the usability evaluation via e-mail invitation. The reported experience level of the participants in \ac{VR} development is divided into 9 novices (8 no and 1 low experience) and 6 experts (3 medium and 3 high experience). All experts are familiar with other \ac{VR} authoring tools.

The usability evaluation was structured in the following way. At first, we sent the participants an archive file containing all required data. The archive file included an introduction to the evaluation process and VREUD, an assignment list, an image file, and a scene as a starting point of the evaluation. After reading the introduction, each participant had to solve a list of assignments. The assignments were divided into the following tasks:

\begin{itemize}
        \item \textbf{Navigation in the Scene:} To evaluate the navigation interface of VREUD, each participant had to navigate to a specific point, select an entity and delete it.
        \item \textbf{Scene Adaptation:} To evaluate the direct manipulation interface of VREUD, each participant had to add an external image. This image had to be manipulated to cover a specific area in the scene by modifying the position, rotation, and scale of the entity. 
        \item \textbf{Single Interaction:} To evaluate the first construction of a simple interaction, each participant had to add an interaction to start a video by pressing a button. The video and button are already contained in the scene. Finally, the participant had to test the interaction in the VREUD web editor.
        \item \textbf{Combined Interactivity:} To further evaluate the interaction design, each participant had to set up a pressure plate, that provides location-based events since it detects when a \ac{VR} user enters and leaves the plate. Afterwards, the participant had to add 5 interactions concerning the interactivity of a presentation. The first interaction enables to open the next page by pressing on a button, the second interaction shows the presentation when the \ac{VR} user enters the pressure plate, the third interaction resets the presentation when entering the plate, the fourth interaction hides the presentation when leaving the pressure plate and the last interaction changes the color of a plate to green when the presentation is completed. Finally, the developed interactivity had to be tested in the VREUD web editor by the participant.
        \item \textbf{Task Construction:} To evaluate the construction of a task, the participant had to create a task. The task contained two activities. The first activity is to go on the stage in the scene. It requires setting up an area. The second activity is to complete the presentation. If the participant has given up on the previous assignment, the participant has to create an activity to look on the fridge in the scene. Finally, the developed task had to be tested in the VREUD web editor by the participant.
\end{itemize}
While performing the assignment, each participant had to measure the time for completing it. The participant was allowed to give up on the assignment which was noted down by them. The participants were not monitored by us, consequently, they had to save after each assignment their session. These save files were used to check if the participant succeeded in the assignment. The succession rate in the combined interactivity assignment is calculated by the successfully designed interactions since the assignment is more complex. After the assignments, the participant had to fill out an online questionnaire that consisted of a \ac{SUS}\cite{SUS} and statements about the difficulty and learnability of the construction of the scene, interactions, and tasks. The statements were rated on a 5-point Likert scale (1 strongly disagree and 5 strongly agree). Finally, the participants had to sent their measured times, their list of given up assignments, and their set of saved sessions to us.

\subsection{Results}
\begin{figure}[htbp]
\centerline{\includegraphics[scale=0.55]{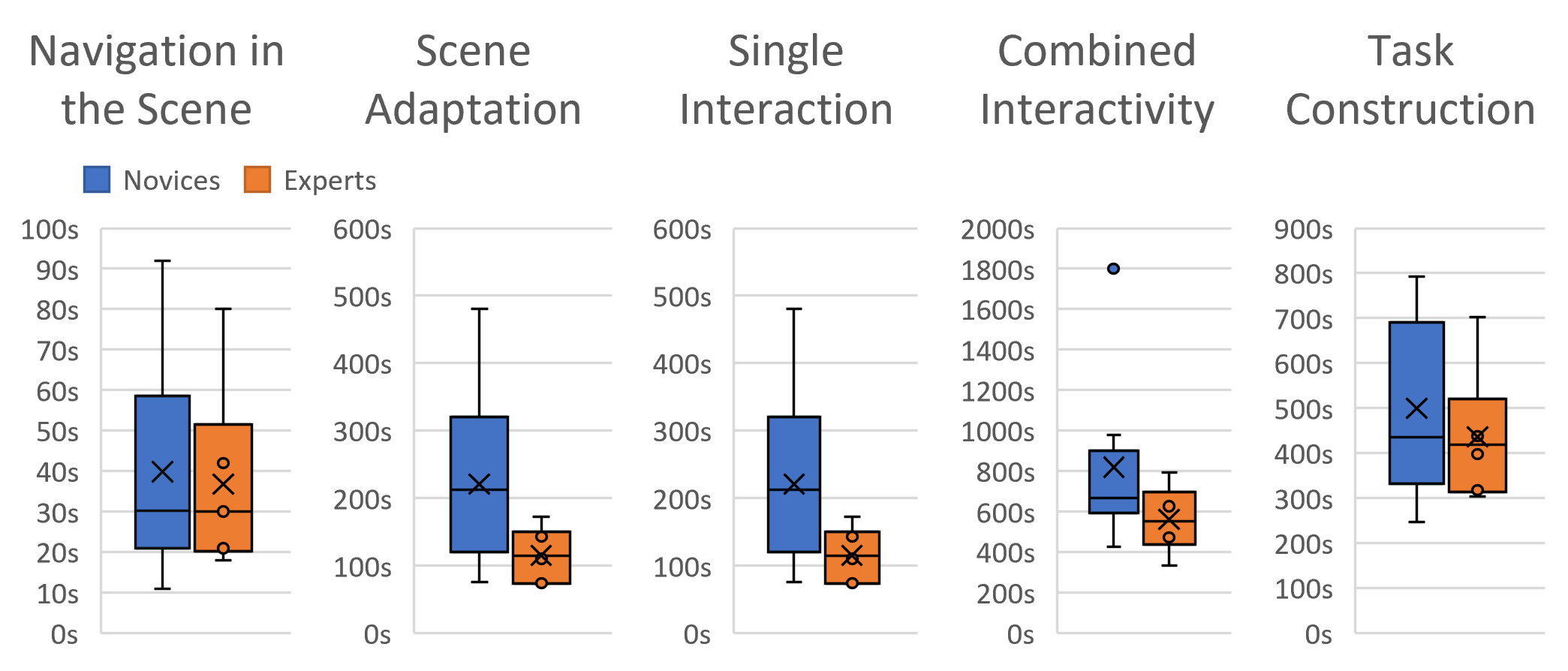}}
\caption{The efficiency results of the usability evaluation.}
\label{efficiency}
\end{figure}

The results concerning efficiency are presented in Fig. \ref{efficiency}. The navigation in the scene was completed in close average times by both groups (37s experts and 40s novices). The scene adaptation was successfully performed on average in 115s by the experts and 206s by the novices. The single interaction construction was succeeded on average in 175s by the experts and 286s by the novices. In the combined interaction, the two groups have a bigger average difference (562s experts and 819s novices). But the median of experts and novices is closer (553s experts and 666s novices). The task construction was completed on average in 437s by the experts and 504s by the novices.

\begin{figure}[htbp]
\centerline{\includegraphics[scale=0.2]{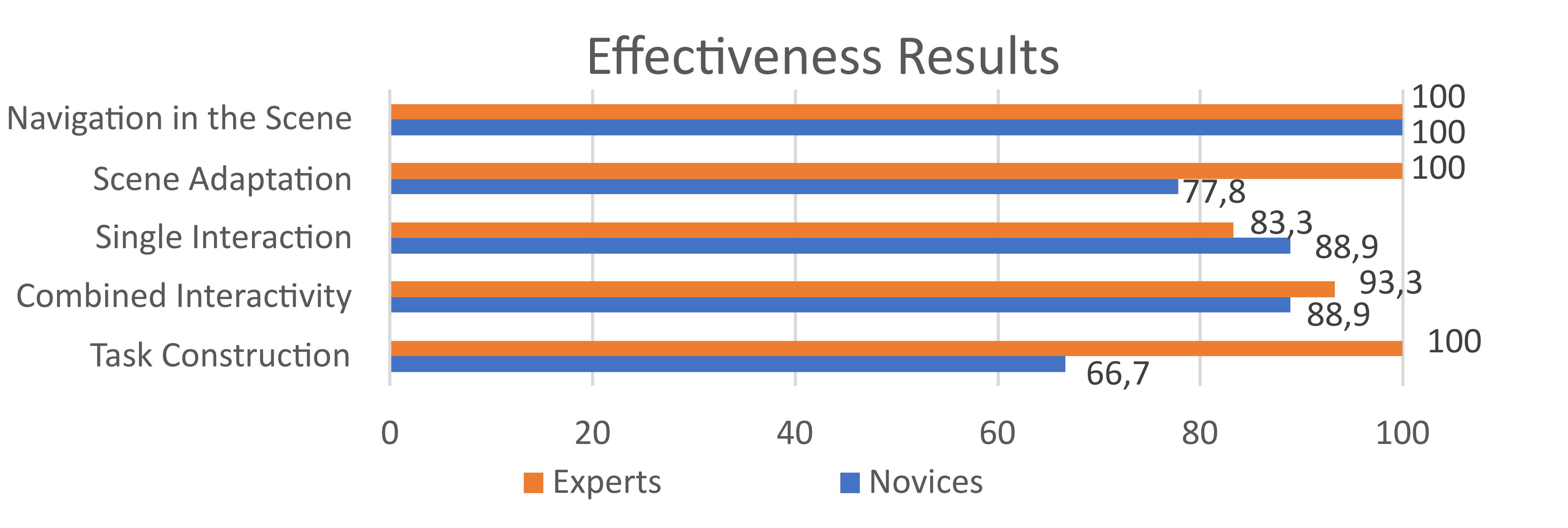}}
\caption{The effectiveness results (in percent) of the usability evaluation.}
\label{effectiveness}
\end{figure}

\begin{figure}[htbp]
\centerline{\includegraphics[scale=0.62]{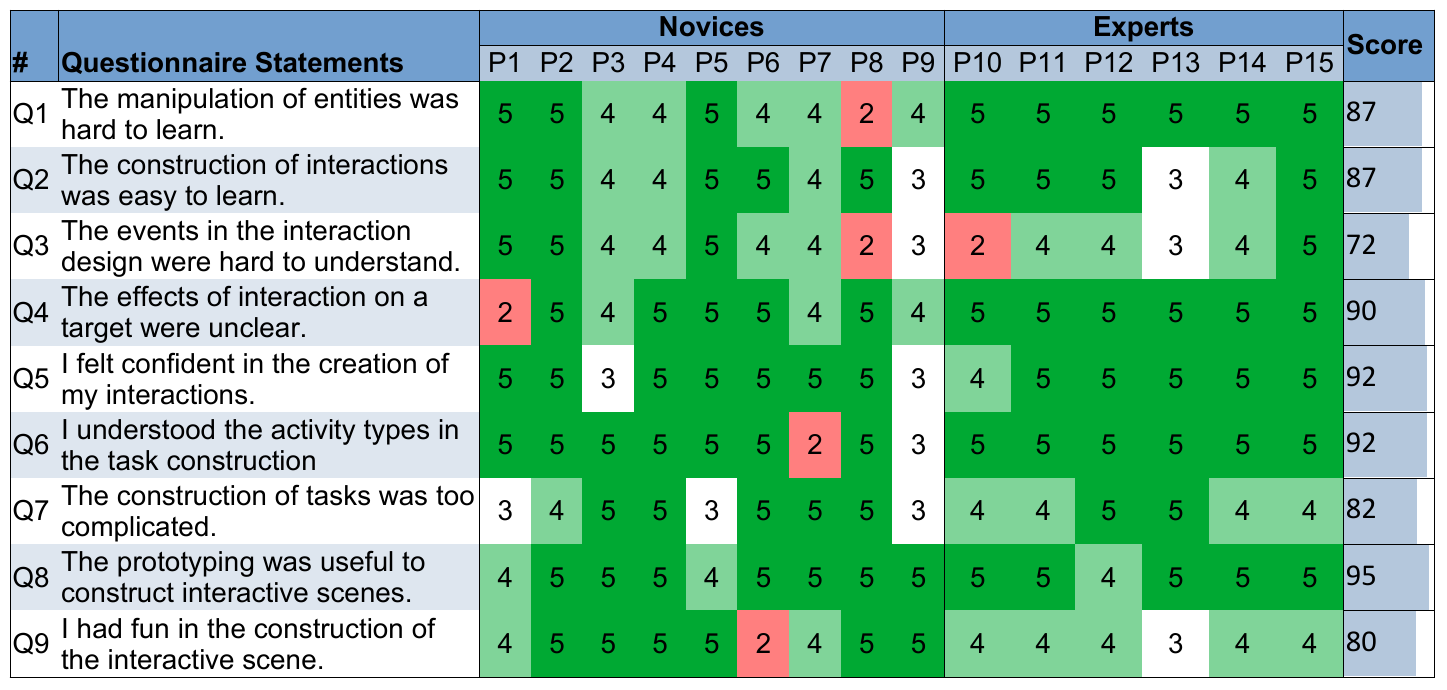}}
\caption{Results of the questionnaire. The results of the negative statements are inverted}
\label{questionnaire}
\end{figure}

The results concerning effectiveness are presented in Fig. \ref{effectiveness}. The navigation in the scene was completed by all participants. The scene adaptation was successfully performed by all experts. Two novices manipulated successfully an image in the scene but they did not add an image. The single interaction construction was succeeded by all except one novice and one expert. In the combined interaction, the novices had constructed two interactions incorrectly and missed adding three interactions. The expert missed adding two interactions. The two participants who failed the previous assignment have succeeded. The task construction was completed by all, except for three novices. Two did not set up the area and one put the area at an incorrect spot.
The results of the questionnaire about the construction of the scene, interactions, and tasks are presented in Fig. \ref{questionnaire}. The highlight of the results is the confidence of the participants to design interactions with VREUD (Q5) which was agreed by all except one novice and one expert. This also affects the results of the effects of the created interactions (Q4) that are clear to the participants and the results of the easy-to-learn statement (Q2) of the interaction construction. The only not so well-performing statement is the understandability of the events in the construction of interactions (Q3) that shows there is room for improvement in the self-explanatory of the events compared to the results of the effects and activity types. However, the participants had still fun in the construction of the interactive VR scene (Q9), which was agreed upon by all except one novice. The best results produced the statement if the prototyping supported the participants in the development of interactive scenes (Q8).

\begin{figure*}[htbp]
\centerline{\includegraphics[width=\textwidth]{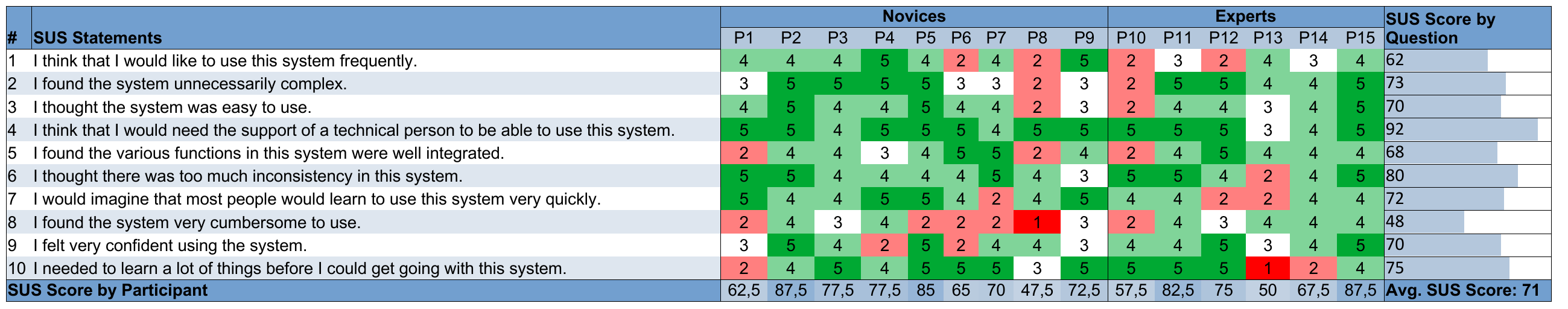}}
\caption{Results of the SUS questionnaire. The results of the negative statements are inverted}
\label{SUS}
\end{figure*}

The results of the \ac{SUS} are presented in Fig. \ref{SUS}. The highlights are the results of the statement if the participants require assistance from a technical person which was disagreed by all except one novice. Another highlight is that the participants agreed on the statement that they did not need to learn much to use VREUD. The statement that VREUD is easy to use was only disagreed by a novice and an expert which is equal to the statement if VREUD is too complicated which was only disagreed by the same participants. The worst results have produced the statement about the cumbersomeness of VREUD. This shows that the usability of the editor can be improved to provide a more intuitive construction. The average SUS score is 71. This is above the mean score of web interfaces (68.2) and close to the average score of the adjective \textit{good} (71.4) \cite{SUS-Score-Int}.

\subsection{Discussion and Limitation}
Our results suggest that VREUD is easy to use and easy to learn. This indicates the potential of VREUD to aid novices in the construction of interactive \ac{VR} scenes since the results of efficiency and effectiveness from both groups are often close together. The user satisfaction also suggests that the component-based and wizard-based development have successfully simplified the construction of interactive \ac{VR} scenes, due to the wide acceptance of the participants. The efficiency results of the single and combined construction of interactions suggest that the novice can design interactions fast after an initial small learning time. This is shown by the fact that the average time of the combined interaction design is not five times higher than the single construction of interaction, although the participant had to create five interactions. The additional setup of the pressure plate indicates that the interaction setup was even faster in the combined interaction construction. However, the results also present that the participants required three to five minutes on average to set up their first interactions. Consequently, the logic embedded in the construction of interactions indicates to require an initial learning time. Combined with the mixed results of the understandability of events, it could affect the results of novices with no technical background. A solution to this issue could be the construction of interactions with the help of interactivity patterns since the higher abstraction suggests that it could help novices who have issues to understand the embedded logic. On the other hand, compared to the great results of the participants in the statement that they felt confident in the construction of interactions, this suggests being a smaller issue. The results of the construction of tasks indicate that a wizard combined with the definition of a task by a set of smaller activities that are configured by an activity type and associated parameters has successfully lowered the complexity of the development. The self-explanatory of the activity types was greatly accepted by the participants. Compared to the logic used in the construction of interactions, the task construction uses generic activities which should be also clear to novices without a technical background, since they are comparable to actions in the real world. For example, an activity type asks the \ac{VR} user to put an object into an area. However, the usability evaluation shows an issue in the current implementation, since three novices have forgotten to set up the area after the construction. Apart from that, the construction of tasks is still well accepted among the participants. Finally, the results of the statement if the participant had fun constructing the interactive \ac{VR} scene suggests that VREUD supplies novices with a playful construction of interactive \ac{VR} scenes, which could motivate them to develop their first interactive VR scene.  

A limitation of the conducted usability evaluation is that VREUD was not evaluated by novices with no technical background, since all participants had a background in computer science. The results could not apply to novices with no technical background. These users could have issues with the embedded logic in the construction of simple interactions. However, the interactivity pattern provides a higher abstraction in the construction and the interactive tutorial supports the novice in their first session. These features combined with the low entry barrier of VREUD could successfully support novices with no technical background in the development.   

\section{Conclusion and Future Work}
In this paper, we introduced a web-based authoring tool for web-based interactive \ac{VR} scenes called VREUD that lowers the entry barrier for novices to develop their interactive \ac{VR} scenes. We achieved this by supporting novices with a component-based construction of the scene, interactions, and tasks of the interactive \ac{VR} scene. We decreased further the complexity by using wizards to focus the user on smaller steps in the development. The effectiveness, efficiency, and user satisfaction of VREUD were evaluated by a conducted usability evaluation. We recruited 15 volunteers with mixed backgrounds in \ac{VR} development to compare the results of novices and experts. Our results show that our tool is easy to learn and that it succeeds in lowering the entry barrier of the development of interactive \ac{VR} scenes.

In future work, VREUD has to be evaluated with a larger group of heterogeneous end-users. With this regard, it should be especially investigated if VREUD supports novices with no technical background in constructing their own interactive VR scenes. The web-based architecture of VREUD and the generated interactive VR scenes animate to use them collaboratively. This arises new challenges in the development, for example, in the construction of social interactions. VREUD could connect desktop users and VR users to support a collaborative design and development of interactive VR scenes. 

\bibliographystyle{./bibliography/IEEEtran}
\bibliography{./bibliography/IEEEabrv,./bibliography/IEEEexample,./bibliography/Paper}
\newpage

\end{document}